\documentclass[prl,aps,amssymb,showpacs,twocolumn]{revtex4}
\usepackage{amsmath}
\usepackage{amssymb}
\usepackage{amsthm}
\usepackage{amsfonts}
\usepackage{enumerate}
\usepackage{latexsym}
\usepackage[dvips]{graphicx}

\newcommand{\beq}{\begin{equation}}
\newcommand{\eneq}{\end{equation}}

\input{epsf}

\begin{document}

\tolerance 10000


\title{The Eight Dimensional Quantum Hall Effect and the Octonions}

\author { Bogdan A. Bernevig$^\dagger$, Jiangping Hu$^\star$,  Nicolaos Toumbas$^\diamond$, and
Shou-Cheng Zhang$^{\dagger}$ }

\affiliation{ $\dagger$ Department of Physics, Stanford
University,
         Stanford, California 94305 \\
         $\star$ Department of Astronomy and Physics, UCLA, Los Angeles, California 90095\\
         $\diamond$ Department of Physics, Harvard University, Cambridge, Massachusetts 02138}
\begin{abstract}
\begin{center}

\parbox{14cm}{We construct a generalization of the quantum Hall
effect where particles move in an eight dimensional space under an
$SO(8)$ gauge field. The underlying mathematics of this particle
liquid is that of the last normed division algebra, the octonions.
Two fundamentally different liquids with distinct configurations
spaces can be constructed, depending on whether the particles
carry spinor or vector $SO(8)$ quantum numbers. One of the liquids
lives on a 20 dimensional manifold of with an internal component
of $SO(7)$ holonomy, whereas the second liquid lives on a 14
dimensional manifold with an internal component of $G_2$
holonomy.}

\end{center}
\end{abstract}

\pacs{03.70.+k, 11.15.-q, 11.25.-w }

\maketitle

The two fundamental mathematical structures (division algebras) a
physicist uses in his everyday life are the real $\mathbb{R}$ and
the complex $\mathbb{C}$ numbers. As we well know, complex numbers
can be treated as pairs of real numbers with a specific
multiplication law. One can however go even further and build two
other sets of numbers, known in mathematics as quaternions
$\mathbb{Q}$ and octonions $\mathbb{O}$. The quaternions, formed
as pairs of complex numbers are non-commutative whereas the
octonions, formed as pairs of quaternionic numbers are both
non-commutative and non-associative. The four sets of numbers are
mathematically known as division algebras. The octonions are the
last divison algebra, no further generalization being consistent
with the laws of mathematics. Strikingly, in physics, some of the
division algebras are realized as fundamental structures of the
Quantum Hall Effect (QHE). The complex $\mathbb{C}$ division
algebra is realized as the fundamental structure of the
two-dimensional QHE \cite{Laughlin}. In a generalization of the
two dimensional QHE, two of us constructed a four-dimensional QH
liquid whose underlying structure is the quaternionic division
algebra $\mathbb{Q}$ \cite{ZhangHu}.

In this paper we present the construction of an eight-dimensional
quantum Hall liquid whose fundamental structure is the division
algebra of octonions. Although this system shares many of the
properties of the previous two dimensional and four dimensional
QHE, its structure is much richer. In particular, our fluid is
composed of particles interacting with an $SO(8)$ background gauge
field. Depending on whether our particles are in the spinor or
vector representation of the $SO(8)$ gauge group, our liquid lives
on a manifold which is either $20$ dimensional or $14$
dimensional. The total configuration space of the liquid is
composed of the eight dimensional base space on which the
particles live plus the configuration space of the particle's
spin. In the four dimensional case, the base space was a four
sphere $S^4$ while the spin space was a two sphere $S^2$ thereby
making the total configuration space six dimensional. The system
reported here is likely to be the last QHE with a rich structure,
and points to connections with the world of string theory. In the
present manuscript we report only the important results of our
analysis, with the full detailed presentation reserved for a
future, longer publication.

The existence of only four division algebras  - $\mathbb{R},
\mathbb{C}, \mathbb{Q}, \mathbb{O}$ -  is related to the existence
of only four (Hopf) fibrations of spheres over sphere spaces: $S^1
\overset{Z_2}{\longrightarrow} S^1$ for $\mathbb{R}$, $S^3
\overset{S^1}{\longrightarrow} S^2$ for $\mathbb{C}$, $S^7
\overset{S^3}{\longrightarrow} S^4$ for $\mathbb{Q}$, and $S^{15}
\overset{S^7}{\longrightarrow} S^8$ for $\mathbb{O}$ \cite{Baez}.
The algebraic structure of the eight dimensional QHE is the
fractionalization of the vector coordinate into two spinor
coordinates fundamental in the Hopf map $S^{15}
\overset{S^7}{\longrightarrow} S^8$:
\begin{equation}
x_a = \Psi^T_\alpha \Gamma^a_{\alpha \alpha '} \Psi_{\alpha '},
\;\;\; \Psi^T_\alpha \Psi_{\alpha } =1, \;\;\; a=1,..., 9
\end{equation}
\noindent Here $\Psi_\alpha$ is a $16$ component real spinor of
$SO(9)$, $\Gamma^a$ is the nine-dimensional real Clifford algebra
of $16 \times 16$ matrices $\{\Gamma^a, \Gamma^b\} = 2\delta_{ab}$
and $x_a$ is a nine component real vector. Due to the
normalization condition, $\Psi$ parametrizes an $S^{15}$.
$\Psi^T_\alpha \Psi_{\alpha } =1$ also gives $x_a x^a = 1$.
Therefore $X_a = Rx_a$ parametrizes a point on the $S^8$ sphere
with radius $R$. An explicit solution of the Hopf map is:
\begin{displaymath}
\Gamma^i = \left(%
\begin{array}{cc}
  0 & \lambda^i \\
  -\lambda^i & 0 \\
\end{array}%
\right), \;\;\; i=1,...,7
\end{displaymath}
\begin{displaymath}
 \Gamma^8 = \left(%
\begin{array}{cc}
  0 & 1_{8 \times 8} \\
  1_{8\times 8} & 0 \\
\end{array}%
\right), \;\;\;
\Gamma^9 = \left(%
\begin{array}{cc}
  1_{8 \times 8} & 0 \\
  0 & -1_{8 \times 8} \\
\end{array}%
\right)
\end{displaymath}
\begin{displaymath}
\left(%
\begin{array}{c}
  \Psi_1 \\
  \vdots \\
  \Psi_8 \\
\end{array}%
\right) = \sqrt{\frac{R+X_9}{2R}} \left(%
\begin{array}{c}
  u_1 \\
  \vdots \\
  u_8 \\
\end{array}%
\right)
\end{displaymath}
\begin{equation}
\left(%
\begin{array}{c}
  \Psi_9 \\
  \vdots \\
  \Psi_{16} \\
\end{array}%
\right) = \frac{1}{\sqrt{2R(R+X_9)}} (X_8 - X_i \lambda^i) \left(%
\begin{array}{c}
  u_1 \\
  \vdots \\
  u_8 \\
\end{array}%
\right)
\end{equation}
\noindent where $\lambda^i$ are seven $8 \times 8$ real antisymmetric
matrices satisfying $\{\lambda^i, \lambda^j\} =-2\delta_{ij}$ which
are constructed from the set of structure constants of the algebra
of octonions \cite{BernevigHuZhang}. The fiber $S^7$ of the Hopf
map is parameterized by the arbitrary eight component real spinor
of $SO(8)$ $(u_1,...,u_8)$ with the normalization condition
$u_\alpha u^\alpha =1$. Any $SO(8)$ rotation on $u_\alpha$
preserves the normalization condition and maps to the same point
on the $S^8$. Since the manifold $S^{15}$ is described by the real
$SO(9)$ spinor $\Psi_\alpha$, the $U(1)$ geometric connection
$\Psi^T d \Psi$ ($U(1)$ Berry phase) is zero, however, the $SO(8)$
connection is non-vanishing $V^T dV = A_a dX_a$, where $V^T=
(\sqrt{\frac{R+X_9}{2R}} I_{8 \times 8},\frac{1}{\sqrt{2R(R+X_9)}}
(X_8 - X_i \lambda^i)^T)$. $A_a$ is the $SO(8)$ gauge field
arising from the connection over $S^8$ and takes the form:
\begin{equation} \label{gaugepotential}
A_\mu  = - \frac{1}{2R(R+X_9)} \Sigma^{\mu \nu} X_\nu, \;\;\; A_9
=0, \;\;
\end{equation}
\noindent where $\mu,\nu =1,...,8$ and where $\Sigma^{\mu \nu}$
are the $28$ generators of the $SO(8)$ Lie algebra: $\Sigma^{\mu
\nu} = (\Sigma^{ij}, \Sigma^{i8})$, $\Sigma^{ij} = -\frac{1}{2}
[\lambda^i, \lambda^j]$, $\Sigma^{i8} = \lambda^i$. Under a
conformal transformation from $S^8$ to Euclidean space $R^8$ this
gauge potential becomes the sourceless, topologically non-trivial
monopole solution of $SO(8)$ Yang-Mills theory of Grossman {\it{et
al}} \cite{Grossman}. The matrices $\Sigma^{\mu \nu}$ are the
$SO(8)$ spin matrices and the gauge potential in
Eq.(\ref{gaugepotential}) can be generalized to any arbitrary
representation of the $SO(8)$ Lie Algebra. The field strength is
$F_{ab} = [D_a, D_b]=  \partial_a A_b - \partial_b A_a +[A_a,A_b]$
where $D_a=\partial_a + A_a$ is the covariant derivative on the
$S^8$ manifold. Component-wise the field strength is $F_{\mu \nu}
= \frac{1}{R^2} (X_\nu A_\mu - X_\mu A_\nu +  \Sigma^{\mu \nu})$
and $F_{9\mu} =- \frac{R+X_9}{R^2} A_\mu$. Regarding the field
strength $F_{\mu \nu}$ as curvature on $S^8$, the Euler number
$\int_{S^8} F \wedge F\wedge F\wedge F$ is $1$, attesting to the
topological non-triviality of our configuration. The field
strength satisfies the generalized self-duality condition $F_{[\mu
\nu} F_{\rho \lambda]} = \epsilon_{\mu \nu \rho \lambda \theta
\chi \beta \gamma} F_{\theta \chi} F_{\beta \gamma}$, where
$\epsilon_{...}$ is the antisymmetric eight dimensional epsilon
symbol.

 We now want to introduce a non-relativistic Hamiltonian for
 particles moving on the $S^8$ in the presence of the $SO(8)$
 gauge field above. The symmetry group of the $S^8$ space is
 $SO(9)$, which is generated by the angular momentum operator
 $L_{ab}^{(0)} = -i(X_a \partial_b - X_b \partial_a)$. Replacing
  the usual derivatives with the covariant derivatives
 which arise due to the presence of the gauge field to get the
 operators $\Lambda_{ab} = -i (X_a D_b - X_b D_a)$,
 the single particle Hamiltonian of particles of mass $M$ yields:
 \begin{equation} \label{hamiltonian}
{\cal{H}} = \frac{\hbar^2}{2MR^2} \sum_{a < b} \Lambda_{ab}^2
\end{equation}
\noindent Similar to the case of the two and four dimensional QHE,
the $\Lambda_{ab}$'s do not satisfy the $SO(9)$ commutation
relations because they are missing the momentum of the monopole
gauge field. When this is added, the newly formed operators
$L_{ab} = \Lambda_{ab} - iR^2 F_{ab}$ satisfy the $SO(9)$
commutation relations and commute with the Hamiltonian
(\ref{hamiltonian}):
\begin{eqnarray}
& &L_{ab} =(L_{\mu \nu}, L_{\mu 9}), \;\;\; \mu,\nu = 1,...,8
\nonumber \\
& & L_{\mu \nu} = L_{\mu \nu}^{(0)} + i \Sigma_{\mu \nu},
\;\;\; L_{\mu 9} = L_{\mu 9}^{(0)} - iRA_\mu \nonumber \\
& & [L_{ab}, L_{cd}] = i [\delta_{ac} L_{bd} + \delta_{bd}L_{ac} -
\delta_{bc} L_{ad} - \delta_{ad} L_{bc}]
\end{eqnarray}
\noindent One can show that $\sum_{a<b} \Lambda_{ab}^2 =
\sum_{a<b} L_{ab}^2 - \sum_{\mu < \nu} \Sigma_{\mu \nu}^2$.

The particles carry $SO(8)$ quantum numbers and the generators
$\Sigma_{\mu \nu}$ combine non-trivially with the orbital angular
momentum $L_{\mu \nu}^{(0)}$ on the equator of $S^8$ to select
only some distinct special representations of the $SO(9)$ $L_{ab}$
which label the particle. Both $SO(9)$ and $SO(8)$ irreducible
representations are labeled by four integers (highest weights)
$(n_1,n_2,n_3,n_4)_{SO(9)}$ and $(n_1,n_2,n_3,n_4)_{SO(8)}$. The
quantum mechanics problem is defined by the choice of the $(n_1,
n_2, n_3, n_4)_{SO(8)}$ representation. Once the monopole
representation is choosen, the composition of the monopole
momentum and the particle orbital momentum on the equator will
specify the $SO(9)$ representation of our particles.

Similarly to the four dimensional QHE, a natural choice is the
monopole in the spinor representation of $SO(8)$
$(0,0,0,I)_{SO(8)}$, where $I$ is an integer. The eigenstates of
the Hamiltonian are specified by the irreducible representations
of $SO(9)$, $(n,0,0,I)_{SO(9)}$ where $n$ is a non-negative
integer specifying the Landau level. The energy eigenvalues are
$E(n,I) = \frac{\hbar^2}{2MR^2} (n^2+nI +7n +2I)$ and the
degeneracy of each state is equal to the dimension of the $SO(9)$
representation $d(n,I) =\frac{1}{10!}
(1+n)(2+n)(3+n)(4+n+I)(1+I)(2+I)(3+I)^2(4+I)(5+n+I)(5+I)(6+n+I)(7+2n+I)$.
The ground state is the lowest $SO(9)$ level for a given $I$ and
is obtained by setting $n=0$, with dimension $d(0,I)$. Therefore
the dimension of the $SO(8)$ representation plays the role of the
magnetic flux whereas $n$ plays the role of the Landau level.

The lowest landau level (lll) wavefunctions are therefore the
$SO(9)$ spinors $(0,0,0,I)_{SO(9)}$. We can obtain these
wavefunctions from the Hopf spinor $\Psi_\alpha$ by observing it
is an eigenstate of the total angular momentum $L_{ab}$:
$L_{ab}\Psi = - \frac{1}{2} \Sigma_{ab} \Psi$. The wavefunctions
can be expanded in the space of the symmetric products of the
fundamental spinor, namely, $Sym\bigotimes^I (0,0,0,1)_{SO(9)}$.
The symmetric product is reducible into $SO(9)$ representations,
with the first representation that it reduces into being
$(0,0,0,I)_{SO(9)}$, which is the only irrep we wish to keep.
Therefore the wave function for the system will be the symmetrized
product along with two conditions that mod out the extra
representations:
\begin{equation}
\Phi=\sum_{\alpha_1,...,\alpha_I}f_{\alpha_1,...,\alpha_I}\Psi_{\alpha_1}...\Psi_{\alpha_I}
\end{equation}
where each $\alpha$ can take values from $0,..., 16$ and where the
coefficients $f_{\alpha_1,...,\alpha_I}$ are symmetric in the
$\alpha$'s and subject to the constraints:
\begin{eqnarray}
\Gamma^a_{\alpha\beta}f_{\alpha\beta\gamma...}=0, \;\;\;
f_{\alpha\alpha\beta...}=0.
\end{eqnarray}
In the thermodynamic limit, both $R, I \longrightarrow \infty$. To
obtain a finite gap, from the dispersion relation we see that the
thermodynamic limit is taken such that $l_0^2= R^2/I$ is kept
constant. $l_0$ can be considered the magnetic length. The
correlation functions can be computed and are gaussian localized.
Our liquid is therefore incompressible.

The degeneracy of each Landau level in the thermodynamic limit $I
\longrightarrow \infty$  varies as $I^{10} \sim R^{20}$. For each
landau level filled, each degeneracy level can carry a particle,
therefore the number of particles in our system varies as
$R^{20}$. Naively, the particle configuration space $V$ is $S^8
\sim R^8$. However, one must keep in mind that the particles carry
spin $\Sigma_{\mu \nu}$ in the spinor representation of $SO(8)$.
The dimension of the configuration space of the spin is naively
equal to the total number of $SO(8)$ generators and varies as
$R^{28}$. However, as said, the spin is an $SO(8)$ spinor and
therefore not the full $SO(8)$ manifold will be available as the
configuration space. In particular, the dimension of the
configuration space available to any representation of a group is
the dimension of the group minus the dimension of the stabilizer
group of that particular representation. The stabilizer of an
$SO(8)$ spinor is $U(4)$, therefore the full configuration space
for our particles will be the number of generators of the coset
$SO(8)/U(4)$, and will vary as $\sim R^{12}$. Hence the total
configuration space is $S^8 \otimes SO(8)/U(4) \approx SO(9)/U(4)$
and the volume scales as $V \sim R^8 \times R^{12} = R^{20}$. Our
fluid has finite density $\rho = N/V$ in the thermodynamic limit.

We now focus on the Newtonian equations of motion derived from the
Hamiltonian ${\cal{H}} + V(X_a)$ where ${\cal{H}}$ is given by
Eq[\ref{hamiltonian}]. We can take the infinite mass limit
$M\rightarrow\infty$ to project the system to the lowest Landau
level. In this limit, the equations of motion are given by
\begin{eqnarray}
\dot{X}_a = \frac{R^4}{I^2} \frac{\partial V}{\partial X_b} F_{ab}
\end{eqnarray}
\noindent Just as in the lll problem, the momentum variables can
be fully eliminated, however, at the price of introducing
non-commuting coordinates. The projected Hamiltonian in the lll is
then simply $V(X_a)$ and the commutation relation is:
\begin{equation} \label{noncomm}
[X_a, X_b] = \frac{R^4}{I^2} F_{ab}
\end{equation}
\noindent We have therefore build a $20$ dimensional Quantum Hall
fluid whose structure in the lll is non-commutative and whose
particles are spinors of $SO(9)$ and interact with an $SO(8)$
gauge field. The configuration space is $S^8 \otimes SO(8)/U(4)
\approx SO(9)/U(4)$. We call this the spinor QH liquid.
Eq(\ref{noncomm}) is very similar to non-commutative relations
appearing in the construction of fuzzy spheres.
\cite{HoandRamgoolam}.

We now obtain a second liquid whose underlying mathematical
structure are the octonions. This is similar to a liquid recently
obtained in the 4DQHE \cite{vectorus}. Let us consider the
Hamiltonian [\ref{hamiltonian}] with the monopole gauge group
$A_\mu$ in the $(I,0,0,0)_{SO(8)}$ representation. This is
realized by picking $\Sigma_{\mu \nu}$ generators of the $SO(8)$
Lie Algebra in the fully symmetric, traceless tensor
representation, which we will call from now on (generalized)
vector. The monopole field again couples to the orbital $SO(8)$
$L_{\mu \nu}^{(0)}$ (but in a different way) to give the general
solution of the Hamiltonian as the $(I+n-m,2m,0,0)_{SO(9)}$
representations of $SO(9)$ where $n, m$ are non negative integers
and $m\leq I$. In order to obtain the vector liquid, we need to
only consider the solutions $(I+n,0,0,0)_{SO(9)}$. We therefore
need an extra projection that will only keep these representations
from the higher space $(I+n-m,2m,0,0)_{SO(9)}$. A elegant way to
impose this projection is to look at the decomposition of
$(I+n-m,2m,0,0)_{SO(9)}$ into $SO(8)$ representations and impose a
physically relevant projection onto the $SO(8)$ space. Attention
should be drawn to the fact that this $SO(8)$ is now the full
orbital plus monopole $SO(8)$, $L^{(0)}_{\mu \nu} + \Sigma_{\mu
\nu}$.

The $SO(8)$ group has the special property that the two (left and
right) spinor representations and the vector representation have
the same dimensions. This property is called triality. It implies
that there are three equivalent $SO(7)$ subgroups of $SO(8)$ which
we will call $SO(7), SO^+(7), SO^-(7)$. The cosets of $SO(8)$ and
its $SO(7)$ subgroups are the seven spheres $S^7, S^{7+}, S^{7-}$
which differ in the form of their metric \cite{WitNicolai} and can
be written as: $SO(8)/SO(7) = S^7$, $SO(8)/SO^+(7) = S^{7-}$, and
$SO(8)/SO^-(7) = S^{7^+}$. Under the $SO(8)$ reduction into
$SO(7)$, the $SO(8)$ vector splits $8 \rightarrow 1+7$ while the
two spinors do not split $8^{\pm}_s \rightarrow 8$. This is the
previous structure, which gives us the 20 dimensional Quantum Hall
spinor liquid. However, under the reduction of $SO(8)$ in
$SO^{\pm} (7)$, the vector of $SO(8)$ gets rotated into the spinor
of $SO^\pm(7)$ and does NOT split, while the spinor of $SO(8)$
rotates and splits into the vectors of $SO^\pm(7)$, $8_s^\pm
\longrightarrow 1+7$. The projection of any antisymmetric tensor
of $SO(8)$ into the $SO^\pm(7)$ subgroups is achieved by the
projection operator:
\begin{equation}
G^\pm_{\alpha \beta \mu \nu} =  \frac{3}{8} (\delta_{\alpha \mu}
\delta_{\beta \nu} - \delta_{\alpha \nu} \delta_{\beta \mu}) \pm
\frac{1}{8} \Omega_{\alpha \beta \mu \nu}
\end{equation}
\noindent where $\Omega_{\alpha \beta \mu \nu}$ is a totally
antisymmetric self dual tensor in 8 dimensions which can be
constructed from the octonionic structure constants
\cite{WitNicolai}. The $(I+n-m,2m,0,0)_{SO(9)}$ decompose into
$\sum_{k_1=0}^{I+n-m} \sum_{k_2=0}^{2m} (I+n-m +k_2 -k_1,
2m-k_2,0,0)_{SO(8)}$. In order to maintain only the vector $SO(9)$
$(I+n,0,0,0)_{SO(9)}$ we need to maintain only the
$\sum_{k_1=0}^{I+n} (I+n -k_1,0,0,0)_{SO(8)}$ representaions of
$SO(8)$ These are the vectors of $SO(8)$. The $SO(8)$ vector is
the only $SO(8)$ representation that transforms in the same way
under projection to both $SO^+(7)$ and $SO^-(7)$. Therefore, the
condition that the projections of a certain representation of
$SO(8)$ into $SO^+(7)$ and $SO^-(7)$ is equivalent guarantees that
we pick up only the vector of $SO(8)$. Projecting the unwanted
representations from the decomposition of $(I+n-m,2m,0,0)_{SO(9)}$
into $SO(8)$ then takes the form of a condition posed on on the
wavefunction: $(G^+_{\alpha \beta \mu \nu} L_{\mu \nu})^2 \Phi
=(G^-_{\alpha \beta \mu \nu} L_{\mu \nu})^2 \Phi$ which reduces
to:
\begin{eqnarray}
\Omega_{\mu\nu\alpha\beta}L_{\mu\nu}L_{\alpha\beta}\Phi =0.
\end{eqnarray}
\noindent The $G^\pm_{\alpha \beta \mu \nu} L_{\mu \nu}$ is the
projection of the $SO(8)$ $L_{\mu \nu}$ into the $SO^\pm(7)$. By
requiring these projections to be equal, it means that they will
be invariant under the maximal subgroup of both $SO^+(7)$ and
$SO^-(7)$. Because $SO^\pm(7) = S^{7\pm} \otimes G_2$, this
subgroup is no other than the automorphism group of octonions,
$G_2$. This can also be seen from the decomposition $SO(8) =
S^{7+} \otimes S^{7-} \otimes G_2$. The discussion above is the
physical interpretation of the mathematical statement that a
vector in eight dimensions is invariant under $G_2$. The vector
here is the octonion, and its automorphism group is $G_2$.

The spectrum of the Hamiltonian with eigenstates
$(I+n,0,0,0)_{SO(9)}$ is given by $E(I,n)= \frac{\hbar^2}{2MR^2}
(n^2+2nI +7n +I)$ and the degeneracy of each state is equal to the
dimension of the $SO(9)$ representation $d(n,I) = \frac{1}{10!}
(1+n+I)(2+n+I)(3+n+I)(4+n+I)(5+n+I)(6+n+I)(7+2n+2I)$. The ground
state (lowest Landau level) is the lowest $SO(9)$ level for a
given $I$ and is obtained by setting $n=0$, with dimension
$d(0,I)$. Higher Landau levels can be obtained by increasing $n$.
In the thermodynamic limit $R,I \longrightarrow \infty$, with a
constant magnetic length $l_0^2 = R^2/I$, there is a finite
constant energy gap between the Landau levels. The degeneracy of
each Landau level in the thermodynamic limit $I \longrightarrow
\infty$ level varies as $I^7 \sim R^{14}$, which is also the
scaling of the number of particles in the level. Hence this vector
liquid is entirely different liquid from the $20$ dimensional
spinor fluid obtained in the case when we pick the monopole field
in the spinor representation of $SO(8)$. The particle
configuration space is $S^8$ multiplied by the configuration space
of the spin $\Sigma_{\mu \nu}$. The spin lives in the vector of
$SO(8)$ and its stabilizer is therefore $SO(7) \otimes U(1)$. The
spin configuration space is therefore $SO(8)/SO(7) \otimes U(1)$,
six dimensional, varying as $R^6$. The full particle configuration
space is therefore $S^8 \otimes SO(8)/SO(7) \otimes U(1) \approx
SO(9)/SO(7)\otimes U(1) \approx S^{15}/U(1)= CP^7 \sim R^{14}$
Therefore $V \sim R^{14}$ and our fluid has finite density $\rho =
N/V$.

The number of particles to fill a Landau level varies as $R^{14}$,
and is globally the manifold $S^{15}/U(1)$ which is the space
$CP^{7}$. In the thermodynamic limit, the wavefunctions of our
system are obtained from the $CP^7$ picture, and can be
constructed from the $\Psi_\alpha$'s by using only half of them.
This is the equivalent of the fact that a complete description of
the lll in the QHE needs use of either the holomorphic or
antiholomorphic functions but not of both. As a coset, $CP^7 =
SU(8)/U(7)$ therefore the wavefunctions will be symmetric tensor
products of $SU(8)$. Writing:
\begin{equation}
\left(%
\begin{array}{c}
 \Phi_1 \\
  \Phi_2 \\
  \Phi_3 \\
  \Phi_4 \\
\end{array}%
\right)= \left(%
\begin{array}{c}
 \Psi_1 + i \Psi_5 \\
  \Psi_2 + i \Psi_6 \\
  \Psi_3 + i \Psi_7 \\
  \Psi_4 + i \Psi_8 \\
\end{array}%
\right);
\left(%
\begin{array}{c}
  \Phi_5 \\
  \Phi_6 \\
  \Phi_7 \\
  \Phi_8 \\
\end{array}%
\right) = \left(%
\begin{array}{c}
  \Psi_9 + i \Psi_{12} \\
  \Psi_{10} + i \Psi_{13} \\
  \Psi_{11} + i \Psi_{15} \\
  \Psi_{12} + i \Psi_{16} \\
\end{array}%
\right)
\end{equation}
\noindent the wavefunctions for the lll on $CP^7$ is $\Phi =
\Phi_1^{m_1} \cdot ... \cdot \Phi_8^{m_8}$  with the constraint
$m_1 + ... +m_8 = I$. $SU(8)$ will act on $\Phi$ while keeping the
subspace $U(7)$ invariant.

We have therefore constructed two eight-dimensional quantum hall
fluids, whose dimensions vary as $R^{20}$ and $R^{14}$. Due to
the invariance under different subgroups of the wavefunction
space, we can say that the $20$ dimensional fluid lives on a space
with $SO(7)$ holonomy, whereas the $14$ dimensional fluid lives on
a space with $G_2$ holonomy.

 We now wish to draw an analogy with M-theory: M-theory lives in
 $11$ dimensions but we are interested in compactifications of
 the theory down to $4$ dimensions that preserve $N=1$ supersymmetry.
 For this reason, we compactify the theory on a 7 dimensional
 compact manifold with $G_2$ holonomy \cite{witten}. The topology of this manifold determines
 much of the structure and content of the effective four dimensional low
 energy theory; for example the amount of supersymmetry and chiral fermions
 that arise from singularities. Similarly in our case, one of the liquids
 lives on a higher $14$ (spatial) dimensional manifold, which is $CP^7=S^{15}/U(1)$.
 This is locally isomorphic to $(S^8 \times S^7)/U(1)$. Locally, we can view
 this as a compactification from $14$ dimensions down to eight dimensions and
 ask what is the structure of the effective $8+1$ dimensional topological
 field theory describing the liquid. The compact space is
 $CP_3=S^{7\pm}/U(1)$
 and it involves a seven sphere with torsion. Because of the torsion the
 holonomy group of the seven sphere reduces to $G_2$. Much of the structure of
the resulting  8+1 dimensional topological effective theory should
be determined by the topology of the compact 7d manifold of $G_2$
holonomy in direct analogy with  M-theory compactifications.

Another interesting feature is related to the field theory of the
QHE presented here. For the 4DQHE the field theory was showed to
be a $U(1)$ Chern Simons on $CP^3$ or a $U(2)$ Chern Simons on
$S^4$ \cite{bernevigtoumbashuchernzhang} . More interestingly, the
four dimensional fluid supports two and four dimensional extended
objects. The 2-branes are the objects which acquire fractional
statistics in the four dimensional quantum hall effect. Similarly,
the vector $14$ dimensional Quantum Hall Effect presented here
will support six and eight dimensional excitations, 6-branes and
8-branes. The 6-branes will have fractional statistics, owing to
the non-vanishing homotopy group $\pi_{15} (S^8) = Z \oplus
Z_{12}$. By dimensional reduction and fuzzification on the fiber
of $S^{15} \overset{S^7}{\longrightarrow} S^8$ we are able to get
to the $S^7 \overset{S^3}{\longrightarrow} S^4$ map and thereafter
to the $S^3 \overset{S^1}{\longrightarrow} S^2$. Hence, the eight
dimensional QHE supports topological excitations which are all the
even branes from zero to eight.

During the development of this work we have learned of a similar
work of S. Thomas and J. Hsu. We hereby acknowledge it.
 We wish to thank D. Bump, S. Thomas, S. Shenker, L. Smolin, B.
Laughlin, S. Kachru, J. Franklin, G. Chapline, Y. Chong, D. Oprea,
C.H Chern, I. Bena and J. Hsu for helpful comments. This work is
supported by the NSF under grant numbers DMR-9814289. BAB is also
supported by the SGF.

\end{document}